\documentclass[aps,prl,twocolumn,showpacs,groupaddress,floatfix]{revtex4}
\usepackage{graphicx}
\usepackage{float}

\begin{document}

\title{Photonic crystal optical waveguides for on-chip Bose-Einstein condensates}
\author{J. Bravo-Abad}
\altaffiliation[Present address: ] {Departamento de F\'{\i}sica Te\'{o}rica de la Materia Condensada
, Universidad Aut\'onoma de Madrid, E-28049 Madrid, Spain.}
\affiliation{Center for Materials Science and Engineering and Department of Physics, MIT, Cambridge, MA 02139}
\author{M. Ibanescu}
\affiliation{Center for Materials Science and Engineering and Department of Physics, MIT, Cambridge, MA 02139}
\author{J.D. Joannopoulos}
\affiliation{Center for Materials Science and Engineering and Department of Physics, MIT, Cambridge, MA 02139}
\author{M. Solja\v{c}i\'{c}}
\affiliation{Center for Materials Science and Engineering and Department of Physics, MIT, Cambridge, MA 02139}

\begin{abstract}

We propose an on-chip optical waveguide for Bose-Einstein condensates based on the evanescent light fields created by surface states of a photonic crystal. It is shown that the modal properties of these surface states can be tailored to confine the condensate at distances from the chip surface significantly longer that those that can be reached by using conventional index-contrast guidance. We numerically demonstrate that by index-guiding the surface states through two parallel waveguides, the atomic cloud can be confined in a two-dimensional trap at about 1$\mu$m above the structure using a power of 0.1mW.
\end{abstract}


\maketitle

% main text

Magnetic microtraps able to guide Bose-Einstein condensates (BECs) have been proposed as a promising way to implement on-chip matter-wave waveguides \cite{Leanhardt02}. However, this technology is restricted by a fundamental limitation: the trapping mechanism works for only a subset of hyperfine states for any given trapping magnetic field. It has been shown that this limitation can be circumvented in a purely optical trap, where atoms in all hyperfine states can be confined \cite{Stamper98}. When working with an optical trap, one has the possibility of modifying the atomic interaction of a BEC by simply varying the external magnetic field \cite{Inouye98}. This could be of paramount importance for the development of efficient Sagnac interferometers based on two counterpropagating matter waves. Due to the mentioned interest, different kinds of dielectric structures have been proposed as optical traps \cite{Renn93,Ito96,Yin98,Barnett00,Arlt01,Burke04,Balykin04,Lien04,Luo04,Rychtarik04}. However, the extension of these ideas to create optical on-chip BEC waveguides and circuits is limited.

One of the simplest schemes one could imagine for on-chip waveguiding of BECs would be to use the evanescent electric field produced by illuminating a high-index waveguide ($\epsilon_{wav}$) deposited on top of a uniform dielectric substrate of dielectric constant $\epsilon_{subs}$ (being $\epsilon_{wav}>\epsilon_{subs}$). In order to have index-guiding in the waveguide, without leakage into the substrate, the wave vector $k$ of the guided mode has to satisfy: $k>(\omega/c)\sqrt{\epsilon_{subs}}$. This in turn means that the decay constant of the evanescent tails into air is given by $\alpha=\sqrt{k^2-\omega^2/c^2}>(2\pi/\lambda_{AIR})\sqrt{\epsilon_{subs}-1}$, where $\lambda_{AIR}$ is the wavelength of light in air. Because $\epsilon_{subs}$ cannot be close to unity (it is typically larger than 2.0), this is a fast decay that would prevent the balance between the repulsion from the surface and the Casimir-Polder attraction. For similar reasons, same constraints apply for any index-guided scheme that involves a solid-state substrate for which  $\epsilon_{subs}-1 \sim 1$. On the other hand, designs without a substrate would have thickness $<\lambda_{AIR}$, rendering them very fragile and difficult to work with. 

\begin{figure}[htb]
\begin{center}
\includegraphics[width=7cm]{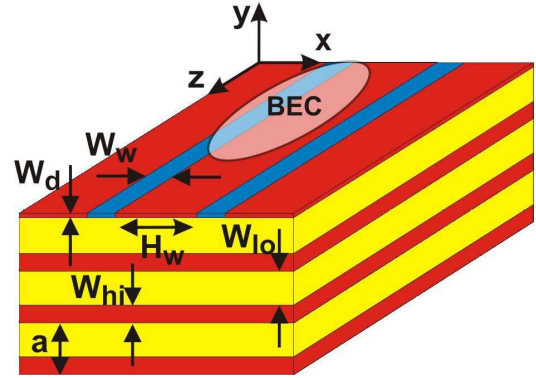}
\end{center}
\caption{Schematic picture of the structure proposed for BEC waveguiding. The system consists of a one-dimensional photonic crystal with a defect layer on its top. In the defect layer two parallel index-guiding waveguides are included. As it is sketched in the figure, for a suitable design of the system, the BEC cloud can be confined in the region between these waveguides. The reference system used throughout the text, together with the geometrical parameters defining the structure are also shown.}
\end{figure}

In this letter, we move a step forward and we show that the use of a photonic crystal (PhC) as a substrate would permit to have decay lengths long enough to circumvent the limits imposed by the Casimir-Polder attraction. We propose a novel design for optical guiding of BECs based on the evanescent light fields created by surface states of one-dimensional photonic crystals\cite{Joannopoulos95} (1D-PhCs). Surface states are prohibited from propagating into the PhC bulk by photonic bandgap effects, and are prevented from leaking into air because of total internal reflection (TIR) effects; thus, they are confined to propagate along the surface. Surface states of this kind are unique to PhCs. 

\begin{figure}[h]
\begin{center}
\includegraphics[width=7cm]{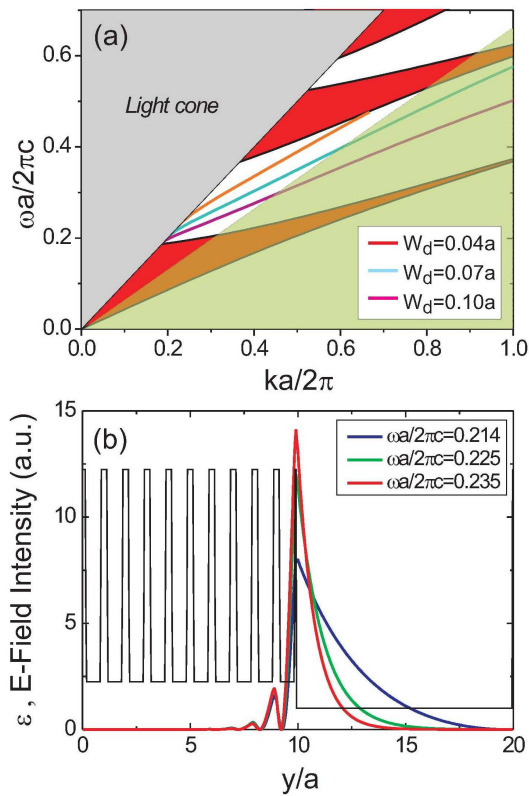}
\end{center}
\caption{(a) Projected band structure of the semi-infinite one-dimensional photonic crystal (PhC) considered in this work [see the profile of dielectric constant in panel (b)]. Red shaded areas show the regions where there exist extended states inside the PhC. Grey shaded area corresponds to eigenmodes that are propagating in air. Orange, blue and magenta lines represent the dispersion relations of the surface states for several values of the thickness of the top layer. Green shaded region corresponds to possible confined states of any defect layer if it lies on top of a uniform dielectric substrate with $\epsilon=\epsilon_{lo}$. (b) Dielectric constant (black line) and electric field intensity profiles along the direction perpendicular to the interface PhC-air. Three surfaces states with frequencies close to the light line are displayed for $W_d$=0.07$a$.}
\end{figure}

Figure 1 shows a schematic picture of the structure analyzed in this work for on-chip BEC waveguiding. The main idea proposed here to get a confining potential both in the x- and y-directions (see the reference system in Fig. 1) is the following. First, we create a 1D-PhC by a periodic sequence of a high dielectric constant ($\epsilon_{hi}$) and a low dielectric constant ($\epsilon_{lo}$) layers (plotted as red and yellow volumes in Fig. 1, respectively). The thickness of each layer is given by $W_{hi}$ and $W_{lo}$. In addition, we define $a$ as the periodicity of the PhC. Secondly, we introduce a defect in this structure by reducing the thickness of its top layer to $W_d$. The electric field of this defect mode is confined both in y$>0$ and y$<0$ regions, i.e., this mode is a surface state. By coupling into the defect layer a laser beam whose frequency is blue-detuned from a certain transition of the atoms forming the considered BEC, the condensate will experience a repulsive potential from the surface. This potential combined with the Casimir-Polder interaction and the force exerted by the gravity (which pushes the atoms towards the surface) will trap the BEC in the minimum of the potential in the y-direction.
 
Once a suitable confinement in the direction normal to the air-PhC interface is obtained, the next step is to design a practical mechanism to confine the BEC in the x direction. We have found that one convenient way to achieve this is to create two index-guided parallel waveguides for these surfaces states, (represented as blue volumes in Fig. 1, where $W_w$ and $H_w$ stand for the width of the waveguides and the separation between them, respectively). 
 
In order to optimize the potential created by the structure described in the previous paragraphs we first analyze the case in which the PhC is uniform in the x-direction (i.e. without waveguides). This is a good starting point because the decay of the electric field in the y-direction will be almost the same as in the system with the two waveguides\cite{Haus84}. 

Red regions in Fig. 2(a) show the projected bands of a semi-infinite 1D-PhC computed by means of the plane-wave expansion method to Maxwell's equations\cite{Johnson01}. We have taken the following values for the parameters defining the structure $\epsilon_{hi}$=12.25, $\epsilon_{lo}$=2.25 and $W_{lo}$=0.7$a$, as they are typical values that could be used in the experiments. In addition we have assumed $\epsilon$=1 in the region above the PhC. The grey shaded region represents modes that can propagate in air. In order to show how the dispersion relation for the surface state changes with the thickness of the defect layer, we have also plotted in Fig. 2(a) the results for three different values of $W_d$ (magenta, blue and orange lines correspond to $W_d$=0.04$a$, $W_d$=0.07$a$ and $W_d$=0.1$a$). Any of these values for $W_d$ could be chosen for our design; for definiteness, from now on we take $W_d$=0.07$a$.

Now, let's turn our attention to study how the electric field profile depends on the frequency of the defect mode. Figure 2(b) shows a cross-section in the y-direction of electric field intensities of the surface states corresponding to three representative frequencies near the light line. Blue, green and red lines render the results for $\omega a/2 \pi c$=0.214, $\omega a/2 \pi c$=0.225 and $\omega a/2 \pi c$=0.235, respectively. The profile of the dielectric constant along the y-direction of the structure has also been plotted. As it is expected, the decay length in air increases as we get closer to the light line (i.e. as the frequency decreases). 

At this point, it is worth taking a closer look at the advantages of the type of systems studied in this work compared with other kinds of rigid dielectric structures that are not based on PhCs. As it is shown in Fig. 2(a), the dispersion relation of the defect mode of a PhC allows us to choose a frequency as close to the light line as we need in our atomic trap design; thereby, the decay length in air can be tuned to be arbitrarily long. However, this property could not be obtained by means of any TIR-based waveguide placed on top of a dielectric substrate, as in that case the frequencies of the confined states  $\omega_d$ are such that $\omega_d < ck/\sqrt{\epsilon_s}$, where $\epsilon_s$ is the dielectric constant of the substrate. This point is illustrated in Fig. 2(a), where the green shaded area corresponds to the region where confined states can exist for the case $\epsilon_s=\epsilon_{lo}$. As we can see in that figure, all of those states are far from the light line, so their decay length in air is short.

Now, before proceeding with the analysis of the full waveguiding structure, let's examine the confining atomic potential created by the 1D-PhC with a defect described previously, but without waveguides. If we illuminate this defect layer with a laser beam whose frequency $\omega$ is blue-detuned with respect to a certain atomic transition of frequency $\omega_0$, then an atom located above the PhC will experience a potential given by the sum of the optical-dipole ($U_{op}$), gravitational ($U_{gv}$), and Casimir-Polder interactions ($U_{CP}$), that is,
$U({\bf r})=U_{op}({\bf r})+U_{gv}({\bf r})+U_{CP}({\bf r})$
The optical dipole contribution can be written as\cite{Grimm00,Balykin00}
\begin{equation}
U_{op}({\bf r})=\beta_s \frac{\hbar \Gamma^2}{8 \Delta}\: \frac{I({\bf r})}{I_{sat}}
\end{equation}                                                                                         
where $\beta_s$ and $I_{sat}$ are two parameters whose value depend on the atomic transition we are considering, $\Gamma$ is the corresponding linewidth and $\Delta=\omega-\omega_0$ (note that in deriving Eq.(1), it has been assumed that $|\Delta| << \omega_0$). $I({\bf r})=\frac{1}{2} \epsilon_0 |{\bf E} ({\bf r})|^2$ is the electric field intensity. Since we are currently assuming that our structure has translational symmetry both in the x and z directions, we can write $I({\bf r})=I_0\:\exp(-2y/\Lambda)$, where $\Lambda$ defines the evanescent decay length of the electric field and $I_0$ is the intensity just above the illuminated layer.

The second term in Eq.(1) represents the gravitational potential and its given by $U_{gv}({\bf r})=mgy$, where $m$ and $g$ stand for the mass of each atom and the gravitational acceleration, respectively. Finally, the last term of Eq.(1) corresponds to the so-called Casimir-Polder potential $U_{CP}({\bf r})=-C_4/y^3$, where the constant $C_4$ is given by \cite{Lin04}
\begin{equation}
C_4=\frac{3 \hbar c \alpha}{32 \pi^2 \epsilon_0} \psi(\epsilon_{hi})
\end{equation} 

In the above expression, $\alpha$ is the atomic polarizability and the value $\psi(\epsilon_{hi})$=0.69 has been computed using the analytical expressions for the function $\psi(\epsilon_{hi})$ given by Yan \emph{et al.} \cite{Yan97,note1}

In order to show the feasibility of the experimental realization of the system presented in this work, we consider the case of Cs atoms, since the confinement of a condensate of Cs atoms in a similar structure has been demonstrated \cite{Rychtarik04}. As we are considering the $D_2$ resonant transition, we must take the following values in Eq.(1), $\beta_s$=2/3, $I_{sat}$=1.1mW/cm$^2$, $\omega_0$=2$\pi\times$3.5$\times$10$^{14}$Hz ($\lambda_0$=857.1nm), and  $\Gamma$=2$\pi \times$5.3$\times$10$^6$ Hz. Regarding the external illumination, we assume that it is produced by a Ti:sapphire laser at $\omega$=2$\pi\times$3.6$\times$10$^{14}$Hz. In addition, to get an optical potential with a similar depth to that obtained in Ref. 13, we suppose that the maximum electric field intensity in the air region very close to the PhC-air interface is $I_0$=2.67$\times$10$^6$ mW/cm$^2$.

\begin{figure}[t]
\begin{center}
\includegraphics[width=\columnwidth]{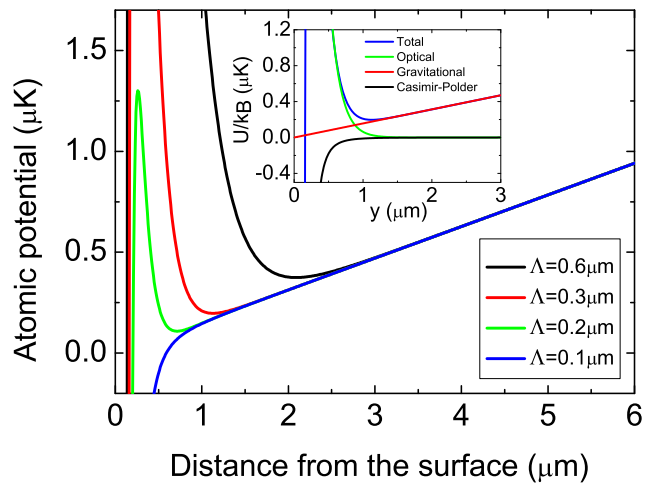}
\end{center}
\caption{Main panel: Total atomic potential for the case of Cs atoms as a function of the distance to the dielectric surface. The results for several decay lengths of the evanescent light field are shown. The optical dipole potential is assumed to be created by a Ti:sapphire laser at a wavelength of 839nm. All the cases plotted in this figure were computed for the same intensity $I_0$=2.67$\times$10$^6$mW/cm$^2$ evaluated just above the PhC-air interface. Inset: Total, optical, gravitational and Casimir-Polder potentials (blue, green, red and black, respectively) for the case of $\Lambda$=0.3$\mu$m. The minimum for the total potential is found at $y_0$=1.1$\mu$m.}
\end{figure}

Figure 3 plots the total atomic potential as a function of the distance from the surface for several values of the decay length $\Lambda$ ranging from 0.1 to 0.6$\mu$m. As it is shown in this figure, a decrease in $\Lambda$ leads to a shift of the minimum of the potential towards the surface. However as can be seen, there exists a minimum distance from the surface where the condensate can be trapped, since if we take $\Lambda<$0.2$\mu$m the total potential does not have a minimum. This limit is established by the fact that in the region where the repulsive optical potential is not negligible for $\Lambda<$0.2$\mu$m, the Casimir-Polder attractive interaction dominates, preventing the formation of a minimum in the total potential. In the case of the corresponding TIR waveguide design with a substrate with $\epsilon=$2.25 one would obtain a maximum decay length $\Lambda<$0.12$\mu$m; thus, given the rapid variation of Casimir-Polder interaction as a function of distance, this demonstrates explicitly that a TIR scheme can not be made to work for the current purpose, and that PhCs indeed have to be used.
 
Since we are interested in bringing the BEC as close as possible to the surface, we will choose $\Lambda$=0.3$\mu$m. This value corresponds to a state of wavevector $ka/2\pi$=0.27 of the defect band, and thereby a value of the periodicity $a$=0.19$\mu$m (see Fig. 2(a)). From now on, we will assume this value of $k$, unless otherwise stated. In that case, the minimum of the potential appears at $y_0$=1.1$\mu$m.

\begin{figure}[t]
\begin{center}
\includegraphics[width=\columnwidth]{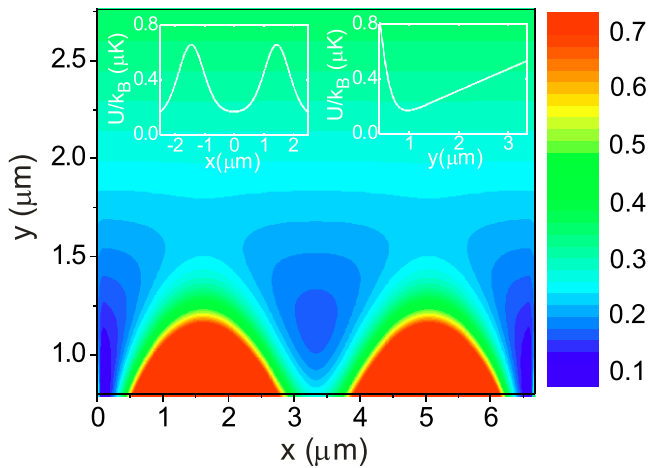}
\end{center}
\caption{2D total atomic potential produced by the structure sketched in Fig. 1 in the xy plane for the case of Cs atoms. Color scale codes the potential in units of $\mu$K. The minimum is found at $(x_0,y_0)$=(3.3,1.1)$\mu$m. Left and right insets show cross-sections of the potential along the lines   $y_0$=1.1$\mu$m and $x_0$=3.3$\mu$m, respectively.}
\end{figure}

It is also interesting to analyze the contribution of each term appearing in Eq.(1) to the total atomic potential. Blue, green, red, and black lines in the insert of Fig. 3 show the total, optical, gravitational and Casimir-Polder potentials, respectively, for $\Lambda$=0.3$\mu$m. This figure reveals that at small distances the Casimir-Polder potential governs the behavior of the total potential while for distances slightly larger that $y_0$, the total potential is dominated by the gravitational interaction.

Finally, let us focus on how to obtain a suitable confining potential in the x-direction, needed to create an optical BEC waveguide. As we described above, we could solve this problem by including in the defect layer two index-guided waveguides of dielectric constant $\epsilon_w$, with  $\epsilon_w > \epsilon_{hi}$ (see schematics in Fig.1).  For definiteness, we take $\epsilon_w$=18 \cite{note2}. The mode we explore is the even mode of this system of two coupled waveguides. In order to reproduce faithfully the features that could be found in the experiments, we have simulated the structure using the Finite-Difference Time-Domain (FDTD) method\cite{Taflove00}. The results obtained with this method come from exact solutions of the 3D Maxwell equations, except for the numerical discretization.

As we described in the introduction, one of the main advantages of our proposal is that there are several parameters that can be changed to our convenience. Through the previous calculations, we have fixed the values of the thickness of the defect layer $W_d$, as well as the wavevector $k$ and the frequency $\omega$ considered. In addition, now for simplicity, we fix the width of the waveguides to $W_w$=2$a$. Therefore, the remaining parameters in our optimization process are the distance between the waveguides $H_w$ and the power $P_0$ carried by the even mode they create. We now look for the values of $(H_w,P_0)$ such as the curvature of the potential at its minimum is similar both in the x- and the y-directions. This condition would be useful if we want to guide a cylindrical-shape BEC along our waveguide. In addition, notice that we require a low value of $P_0$ in order to have an efficient guiding system.

Following the program described above, we have found a suitable configuration for $H_w$=3.2$\mu$m and $P_0$=0.1mW. Figure 4 shows a contour plot of the corresponding two dimensional total atomic potential in the xy plane. The color scale codes the potential in units of $\mu K$. This figure shows a minimum at $(x_0,y_0)=(3.3,1.1)\mu$m, being the center of the waveguides located at $x_1$=1.5$\mu$m and $x_2$=5.1$\mu$m. Left and right insets plot cross-sections along $y_0$=1.1$\mu$m and $x_0$=3.3$\mu$m, respectively. From these cross-sections, we find that the trap frequency at the minimum along the x-axis is $\omega_x$=1233Hz, while along the y-axis it is $\omega_y$=642Hz. This corresponds to a ratio  $\omega_x/\omega_y$=1.92, which confirms that it is possible to get a confining potential with similar curvatures in both transverse directions using this configuration. Notice that, with the same power, we could get other shapes for the BEC by changing the ratio $\omega_x/\omega_y$ through the variation of $H_w$.

In conclusion, we have proposed a new class of optical BEC waveguides based on illuminating two waveguides built into a defect of a 1D-PhC. We have demonstrated that a power as low as 0.1mW is enough to guide a BEC of Cs atoms through the suggested waveguide at about 1$\mu$m above the structure. The kind of system proposed in this work could be considered as a basic ingredient for the development of novel BEC on-chip devices.

We are thankful to Wolfgang Ketterle for providing inspiration and suggestions for this work. J.B.A. acknowledges financial support by the Spanish MEC under grant BES-2003-0374.


\begin{thebibliography}{}
\bibitem{note1} Although the calculation of the Casimir-Polder attraction of the full multilayer structure would require a more detailed analysis, by assuming   we obtain an upper limit for the strength of this interaction.
\bibitem{note2} Equivalently, bearing in mind a possible experimental implementation, one could choose a smaller value of $\epsilon_w$ and a larger local thickness $W_d (x)$ of the defect layer.
\end{thebibliography}
\end{document}